\RequirePackage{lineno}
\documentclass[preprint,12pt,nofootinbib]{elsarticle}

\usepackage{graphicx,amssymb}

\usepackage{color}

\makeatletter
\AtBeginDocument{\@ifpackageloaded{natbib}{\ifNAT@numbers\if@filesw\immediate\write\@auxout{\string\global\string\NAT@numberstrue}\fi\fi}{}}
\makeatother
\usepackage{amstext}

\begin{document}

\setpagewiselinenumbers
\modulolinenumbers[1]
\begin{frontmatter}

\title{A simulation-based study of the neutron backgrounds for NaI dark matter experiments}

\author[ibs]{E.J.~Jeon}
\author[ibs]{Y.D.~Kim}

\address[ibs] {Center for Underground Physics, Institute for Basic
  Science (IBS), Daejon 305-811, Korea}

\begin{abstract}
Among the direct search experiments for weakly interacting massive particle~(WIMP) dark matter, 
the DAMA experiment observed an annual modulation signal  
interpreted as WIMP interactions with a significance of 9.2$\sigma$. 
Recently, Jonathan Davis claimed that the DAMA modulation may be
interpreted on the basis of the neutron scattering events induced by the muons and
neutrinos together. We tried to simulate the neutron backgrounds at
the Gran Sasso and Yangyang laboratory with and without the
polyethylene shielding to quantify the effects of the ambient neutrons
on the direct detection experiments based on the crystals. 
\end{abstract}

\end{frontmatter}



\section{Introduction}

Numerous astronomical observations have led to the conclusion that the majority
of the matter in our universe is invisible, exotic, and
nonrelativistic dark matter \cite{Komatsu:2010fb,Ade:2013zuv}.
However, it is still unknown what the dark matter is. 
Weakly interacting massive particles~(WIMPs) are one of the most attractive
dark matter particle candidates \cite{lee77,jungman96}. 
The lightest supersymmetric particle
(LSP) is a possible WIMP candidate theoretically hypothesized beyond the standard model
of  particle physics. There have been a number of experiments that directly search for
WIMPs in our galaxy by looking for WIMP--nucleus scattering through nuclear recoil~\cite{gaitskell04,baudis12}. 

To date,  several experiments (e.g.,  DAMA \cite{bernabei08,bernabei10}, CoGeNT \cite{aalseth11,Aalseth:2012if}, 
CRESST \cite{cresst730kg}, and CDMS \cite{agnese13}) have yielded signals
that could be interpreted as being possibly
due to WIMP interactions.
Except for the DAMA experiment, the
indications of WIMP-like signals are not significant at the \textless3$\sigma$ confidence level, 
and the same experimental groups or independent groups are planning to confirm these signals.
However, the result from the DAMA experiment has attracted attention
because the observation of an annual modulation of
WIMP-like signals with a
significance of 9.2$\sigma$ has been reported. This finding has
spurred a continuing debate concerning the observation of WIMPs 
over the past 15 years.
The WIMP--nucleon cross sections inferred
from the DAMA modulation are in conflict with limits from other
experiments that directly measure the nuclear recoil signals, such as
XENON100~\cite{aprile12}, LUX~\cite{agnese14}, and SuperCDMS~\cite{akerib14}. 
However, it is possible to explain all these experimental results without conflict
because of nontrivial systematic differences in detector responses~\cite{dsys1,dsys2} and the commonly used astronomical model
for the WIMP distribution~\cite{cmodel}. 

Recently Jonathan Davis of Durham University has proposed a new model for the DAMA annual modulation, which is
a sum of two annually modulating components with different
phases. More specifically, the events are composed of neutrons, which
are liberated in the material surrounding the detector by a
combination of 8B solar neutrinos and atmospheric muons~\cite{davis2014}.
The DAMA group 
disagreed with Davis's claim because the induced modulation
amplitudes from neutrons induced by muons and by neutrinos are less
than  9 $\times 10^{-6}$ cpd/kg/keV (2 $\times 10^{-5}$ cpd/kg/keV for
neutrons produced in the lead shield) and less than 2 $\times 10^{-��6}$
cpd/kg/keV, respectively from the simulation, and they are less than  0.1\% of the 
modulation amplitude measured by DAMA/LIBRA \cite{bernabei2014}. 
Further, Barbeau et al criticized Davis's claim because 
a seven order of magnitude discrepancy in the neutron contribution was required \cite{barbeau2014}.

\section{Method of simulation}
To understand the neutron backgrounds for NaI dark matter experiments, 
we have performed simulations with the GEANT4 Toolkit~\cite{geant4}.
The hadronic models of muon nucleus interaction and muon decay at rest were used in the GEANT4 simulation, 
version of 9.6.p02~
(G4MuonVDNuclearModel and G4MuonMinusCaptureAtRest). 

We have simulated the cosmic muons and the neutrons induced 
by simulating muons passing through the rock directly. We did not simulate the scintillation 
and light collection processes inside the crystals 
because it is not required for the conclusion derived in this study.
Further, we did not simulate the environment neutrons underground because presumably these neutrons
will not have annual modulation signals.

\section{Analysis}

\subsection{Muon energy spectrum}
Neutrons are produced primarily by the muons in the rock and materials in the detection system. 
These neutrons generate secondary neutrons via hadronic interactions in the materials.
We simulated all the primary and secondary processes using GEANT4 simulation program starting with muons passing through the rock.

We used the muon energy spectrum provided by D.~M.~Mei and A.~Hime~\cite{mei}, which is given by

\begin{equation}
  \frac{dN}{dE_{\mu}} = Ae^{-bh(\gamma_{\mu}-1)} \cdot (E_{\mu} + \epsilon_{\mu}(1-e^{-bh}))^{-\gamma_{\mu}}
\end{equation}

where h is the rock slant depth in km.w.e and we used a set of parameters provided by Groom \emph{et~al.} for $\epsilon_{\mu}$, b, and $\gamma_{\mu}$.

\begin{figure}[!htb]
\begin{center}
\includegraphics[width=0.6\columnwidth]{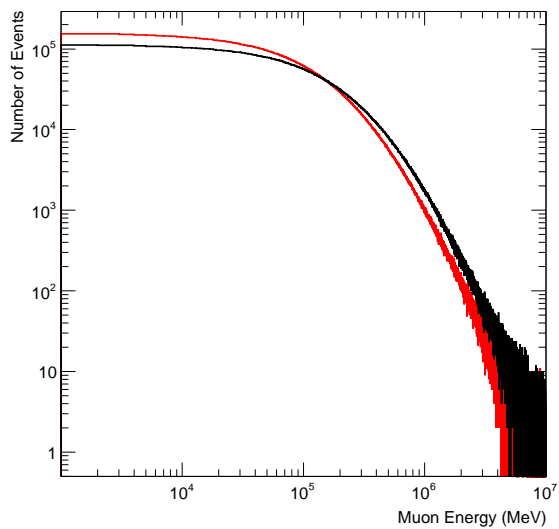}
\caption{Muon energy spectrum at Gran Sasso (blue) and at Yangyang (red).}
\label{muonsp}
\end{center}
\end{figure}

For two different depths at Gran Sasso and at Yangyang laboratory~(Y2L) we assumed that h=3.1~km.w.e and h=1.8~km.w.e, respectively.
Figure~\ref{muonsp} shows muon energy spectra generated at Gran Sasso~(black line) and at Y2L~(red line).
Muon average energies are 276~GeV and 201~GeV for Gran Sasso and Y2L, which are consistent with the measured values~\cite{mei}.

\subsection{Schematic layout of simulation geometry}
\begin{figure}[!htb]
\begin{center}
\includegraphics[width=0.6\columnwidth]{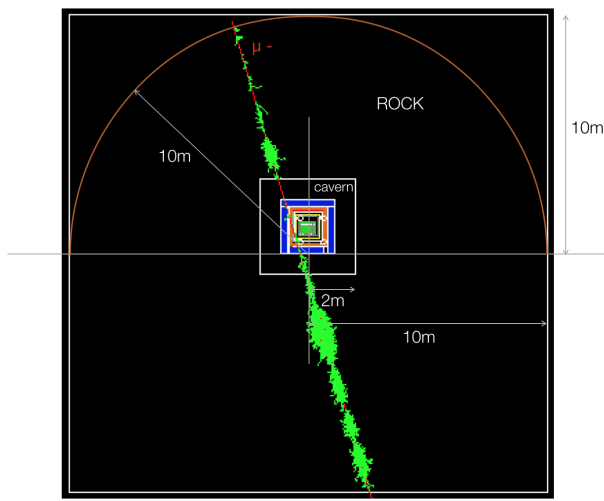}
\caption{Schematic layout of simulation geometry.} 
\label{layout}
\end{center}
\end{figure}

Inside the rock of a thickness of $20~m\times20~m\times20~m$ 
there is an air-filled cavern whose dimensions are $4~m\times4~m\times4~m$. 
A hemisphere with a radius of 10 m is assumed and 
generated muons are incident onto the shielded detector with an angle randomly selected according
to an angular distribution proportional to $cos^{2}\theta$, as shown in Figure~\ref{layout}. The vertex of the
muons is generated uniformly inside the square of 20~m$\times$20~m
area at the surface of the hemisphere. The size of the square is
determined to cover the whole volume of the rock for all the
angles of the generated muons. Figure~\ref{muonsp} shows the energy spectrum
of generated muons at the hemisphere.
The muon flux at Gran Sasso is considered as $2.7\times10^{-8}~muons/cm^{2}/s$ 
or $2\times10^{7}~muons/(20~m\times20~m)/2144~days$ and it is $2.7\times10^{-7}~muons/cm^{2}/s$ or 
$2\times10^{7}~muons/(20~m\times20~m)/214~days$ at Y2L.

\subsection{Neutron production at the boundary of rock and cavern}
\label{sec3.3}
\begin{figure}[!htb]
\begin{center}
\includegraphics[width=0.6\columnwidth]{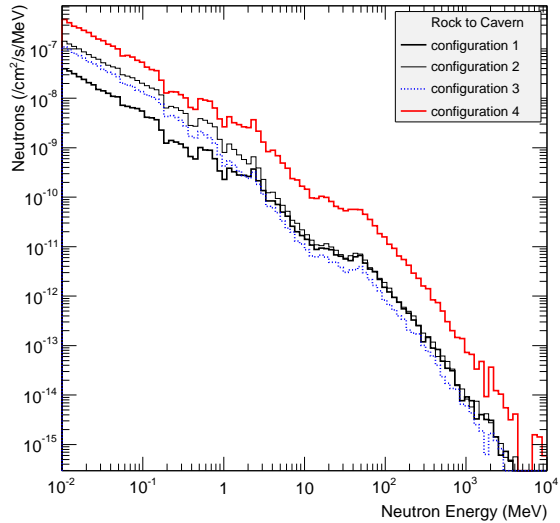}
\caption{Neutron energy spectra at the boundary between the rock and cavern with different shielding configurations inside the cavern. In the simulation, there are four shielding configurations: 
configuration 1~(thick black line), configuration 2~(thin black line), configuration 3~(dotted blue line), and configuration 4~(thick red line).}
\label{neutronsp}
\end{center}
\end{figure}

To validate the simulation for the neutron production 
we compared the neutron fluxes entering the cavern, which are produced in the rock. 
We also included neutrons scattered back from the rock wall as well as shielding materials surrounding the detector and reentering the cavern. 
Figure~\ref{neutronsp} shows the neutron energy spectra at the boundary between the rock and cavern, 
obtained from the simulation of muon propagation and interaction with materials in the rock. 
There are four different shielding configurations inside the cavern, denoted as 1, 2, 3, and 4, defined in section~\ref{sec3.4}, 
therefore, we can see the effect of the neutrons reflected back from the wall and shielding materials inside the cavern. 
Most are dominant below $\sim$1~MeV and it depends on the outmost shield layer whether it is polyethylene~(PE) or not.

According to the simulation results neutron fluxes are $6.98\times 10^{-10}~n/cm^{2}/s$ and $5.3\times 10^{-9}~n/cm^{2}/s$, for energies above 1~MeV,  
under shielding configuration 1 at Gran Sasso and under configuration 4 at Y2L, respectively. 
It is in good agreement with GEANT4 results of version 6.2~\cite{HM} and close to the FLUKA result~\cite{fluka}, $8.7\times 10^{-10}~n/cm^{2}/s$,  
obtained from muon simulation in NaCl at Boulby.

\subsection{Shielding configurations and shielding effects}
\label{sec3.4}
We assumed the NaI detectors used by DAMA/LIBRA, and therefore 
there are 5$\times$5 segmented NaI crystals with dimensions of $(10.2\times 10.2\times 25.4)~cm^{3}$ for a single crystal.
The target NaI crystals are inside 
a 10~cm copper shield and it is additionally surrounded by lead and polyethylene~(PE) shields.

To study the shielding effects, we set four different shielding configurations outside the copper shield:

\begin{enumerate}
  \item Lead (15~cm) + PE (30~cm) same as the DAMA setup
  \item PE (30~cm) + Lead (15~cm)
  \item Lead shield only (15~cm)
  \item PE (5~cm) + Lead (15~cm) + MD (30~cm) same as KIMS-NaI setup
\end{enumerate}

\begin{figure}[!htb]
\includegraphics[width=0.52\columnwidth]{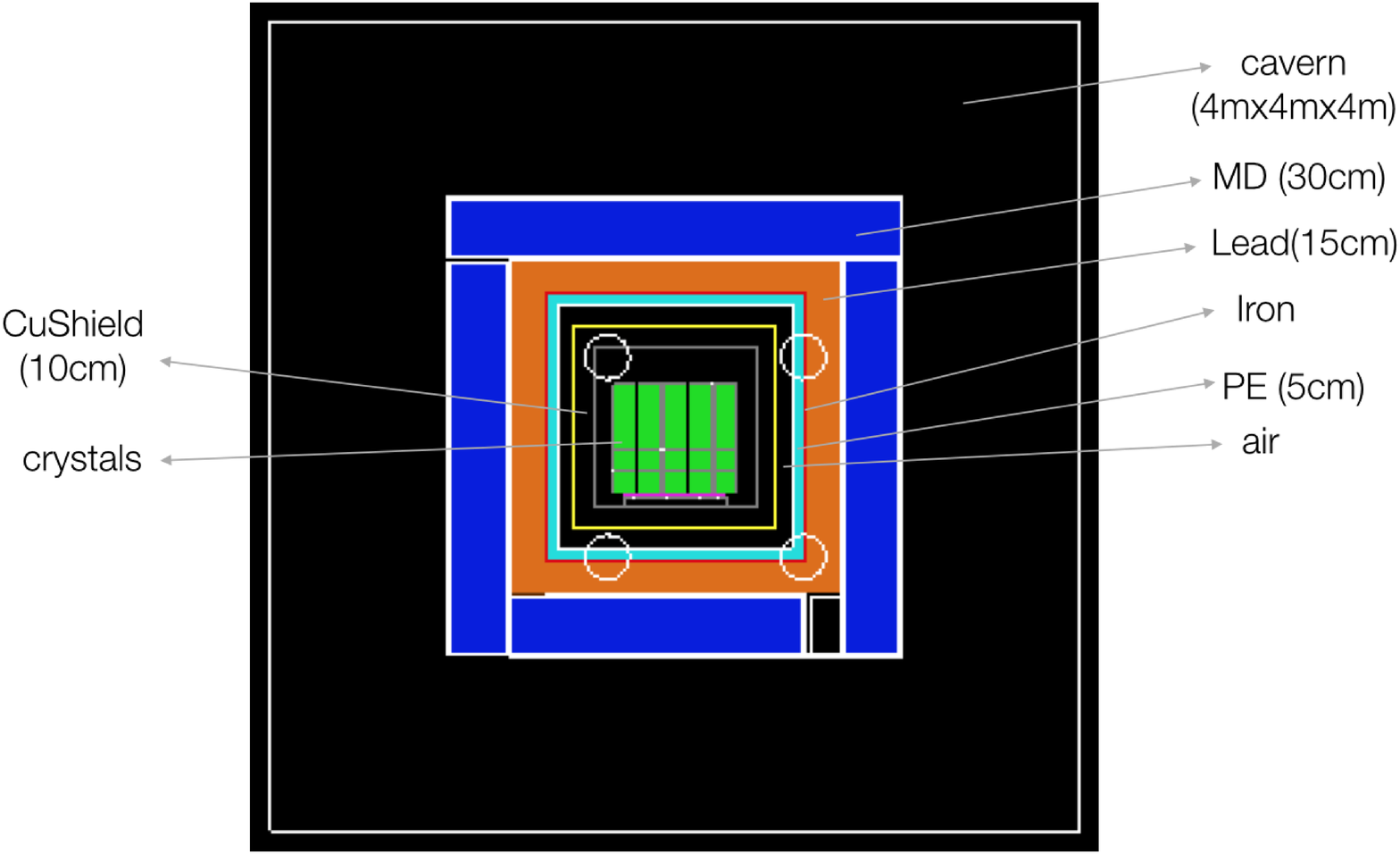}
\includegraphics[width=0.52\columnwidth]{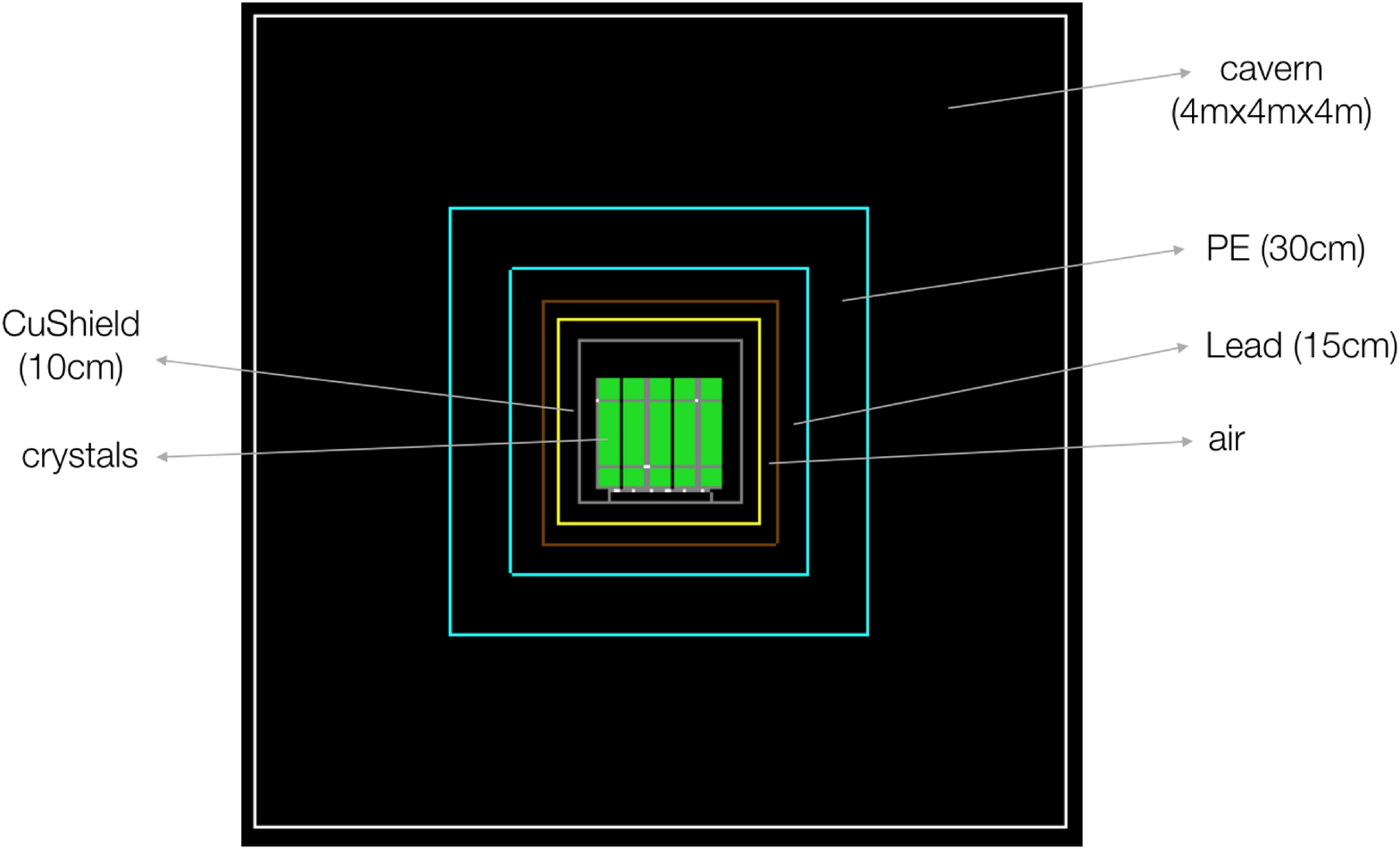}
\caption{Shielded detector geometries for KIMS-NaI (left) and DAMA (right).}
\label{shield}
\end{figure}

Shielding configuration 1 represents the DAMA setup at Gran Sasso depth, which uses 30~cm thick PE as the outmost shield layer. 
The order of lead and PE shield layers was switched for configuration 2.
The shielding effect was also tested with a lead shield only~(configuration 3).
Figure~\ref{shield} shows shielded detector geometries for KIMS-NaI~(left) and DAMA~(right) setups. 
For the KIMS-NaI setup~(configuration 4), 
5~cm thick PE was put inside a lead shield and MD outside the lead shield, which is a muon detector filled with 
mineral oil (liquid paraffin). We expect that 30~cm thick MD can also shield neutrons as much as the PE shield did in the configuration 1. 
 
\begin{figure}[!htb]
\begin{center}
\includegraphics[width=0.49\columnwidth]{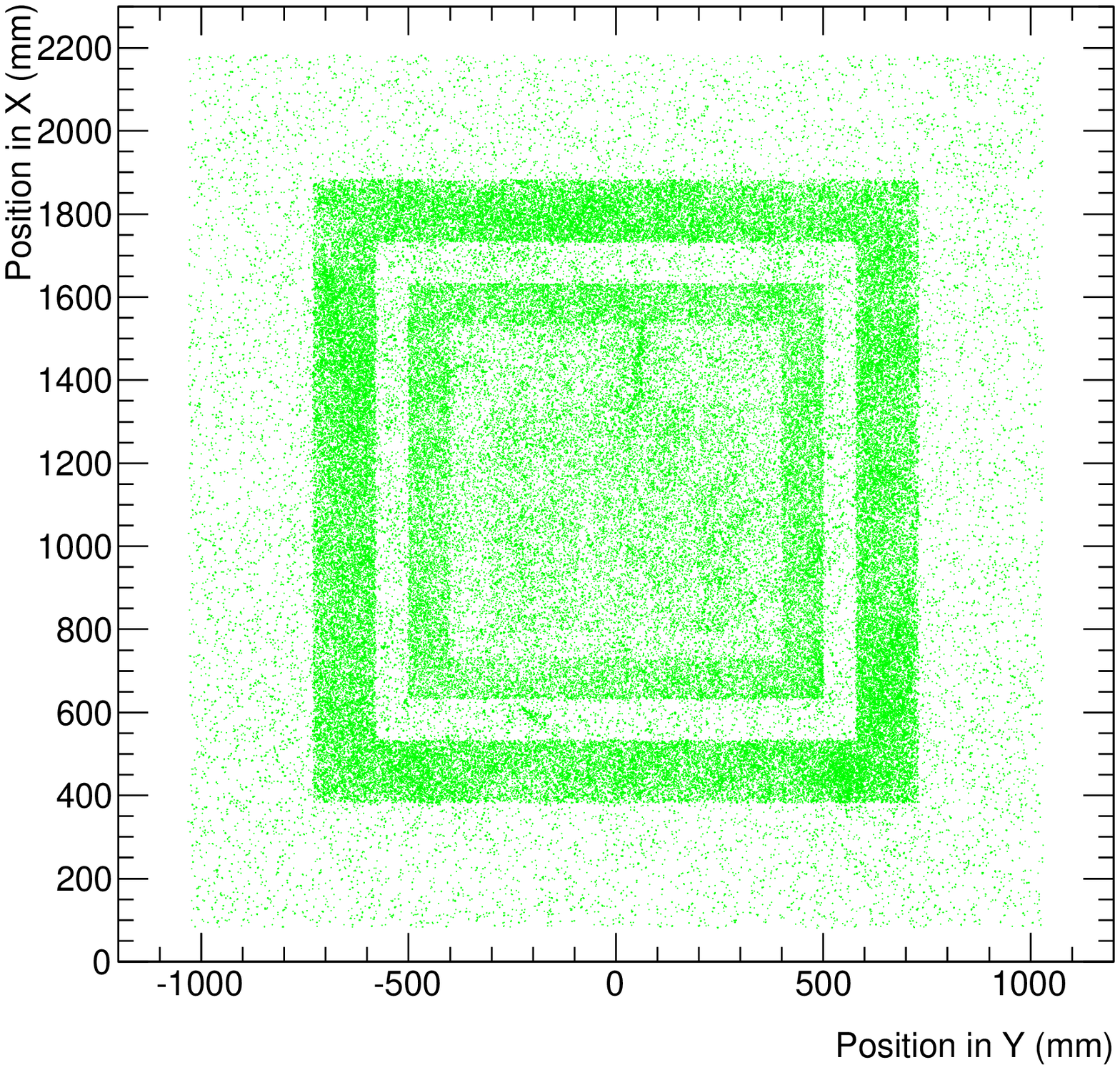}
\includegraphics[width=0.49\columnwidth]{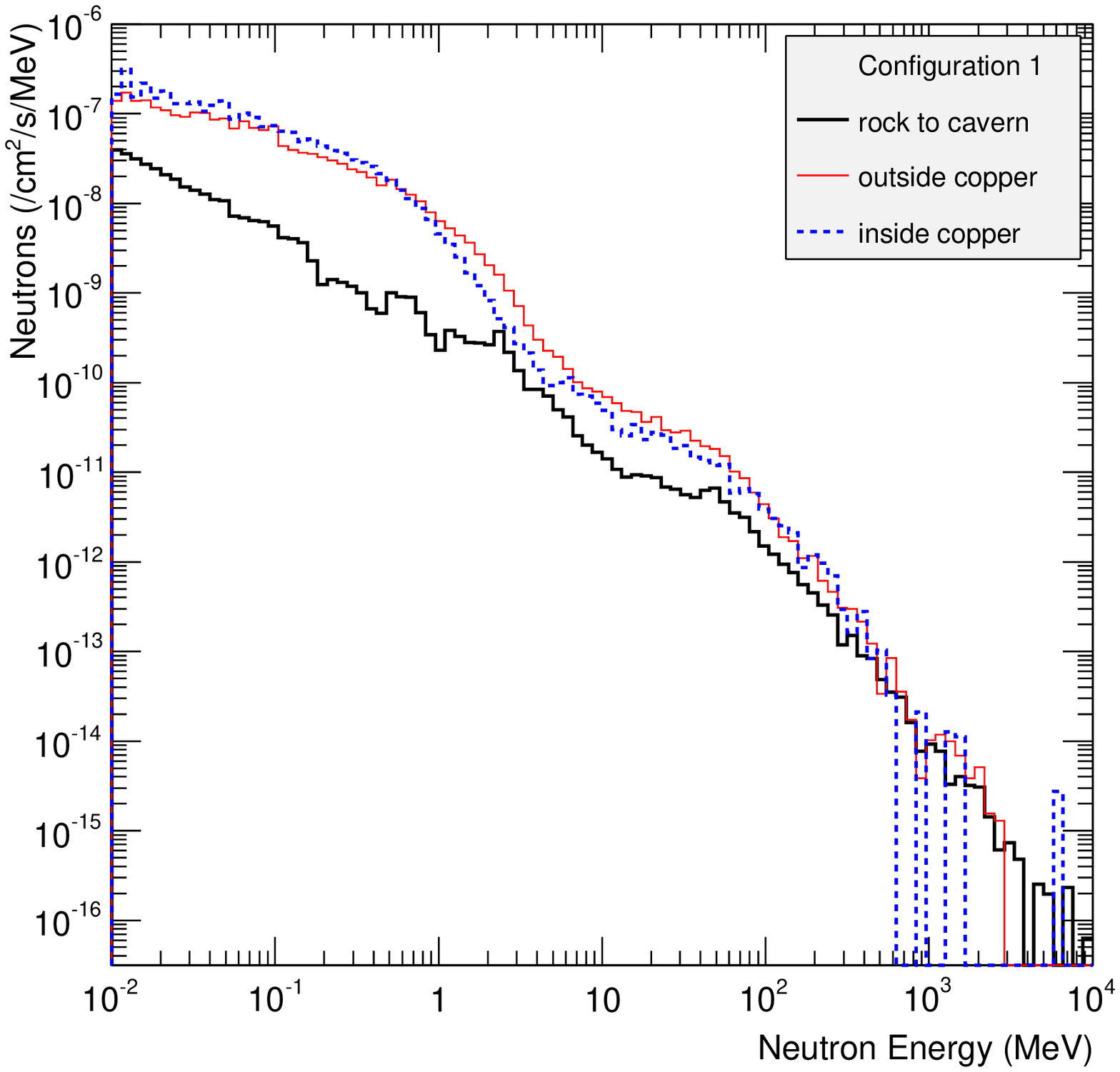}
\caption{Neutron vertex distribution (left) and neutron energy spectra at boundaries between different shield layers (right).}
\label{vertexAndNeutronBoundaryGeom1}
\end{center}
\end{figure}

However, neutrons can also be produced by muon interaction with shielding materials and should be considered in configuring shield layers. 
Figure~\ref{vertexAndNeutronBoundaryGeom1}~(left) shows the primary vertex positions of neutrons in the x-y plane of the shielding configuration 1, 
and it is found to be dense in the lead and copper shields. 
The right side of Figure~\ref{vertexAndNeutronBoundaryGeom1} 
shows neutron energy spectra at boundaries between different shield layers.
We defined three different boundaries: rock/cavern~(thick black line), outside copper shield~(thin red line), 
and between copper and target crystals inside~(dotted blue line).  
From the neutron spectra we found that the neutrons are built up in the lead and copper shields. 
 
\begin{figure}[!htb]
\begin{center}
\includegraphics[width=0.5\columnwidth]{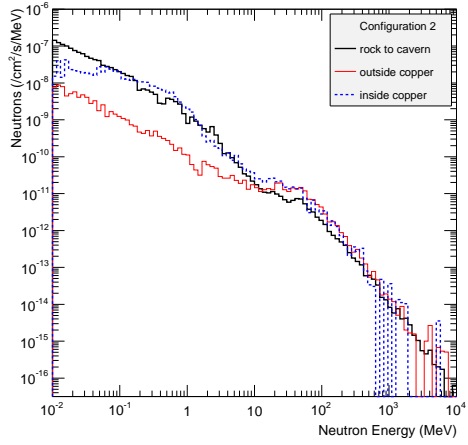}
\caption{Neutron energy spectra at boundaries between different shield layers with configuration 2.}
\label{neutronsAtBoundariesGeom2}
\end{center}
\end{figure}

To quantify the effects of neutrons with and without PE inside the lead shield 
we tested neutron energy spectra at boundaries with configuration 2, in which we changed the order of shield layers of PE and lead, as explained earlier.
As shown in Figure~\ref{neutronsAtBoundariesGeom2} neutrons below $\sim$10~MeV are reduced dramatically between PE and the copper shield. 
Therefore, it is more effective to shield neutrons below $\sim$10~MeV by putting the PE shield inside the lead.

\begin{figure}[!htb]
\begin{center}
\includegraphics[width=0.49\columnwidth]{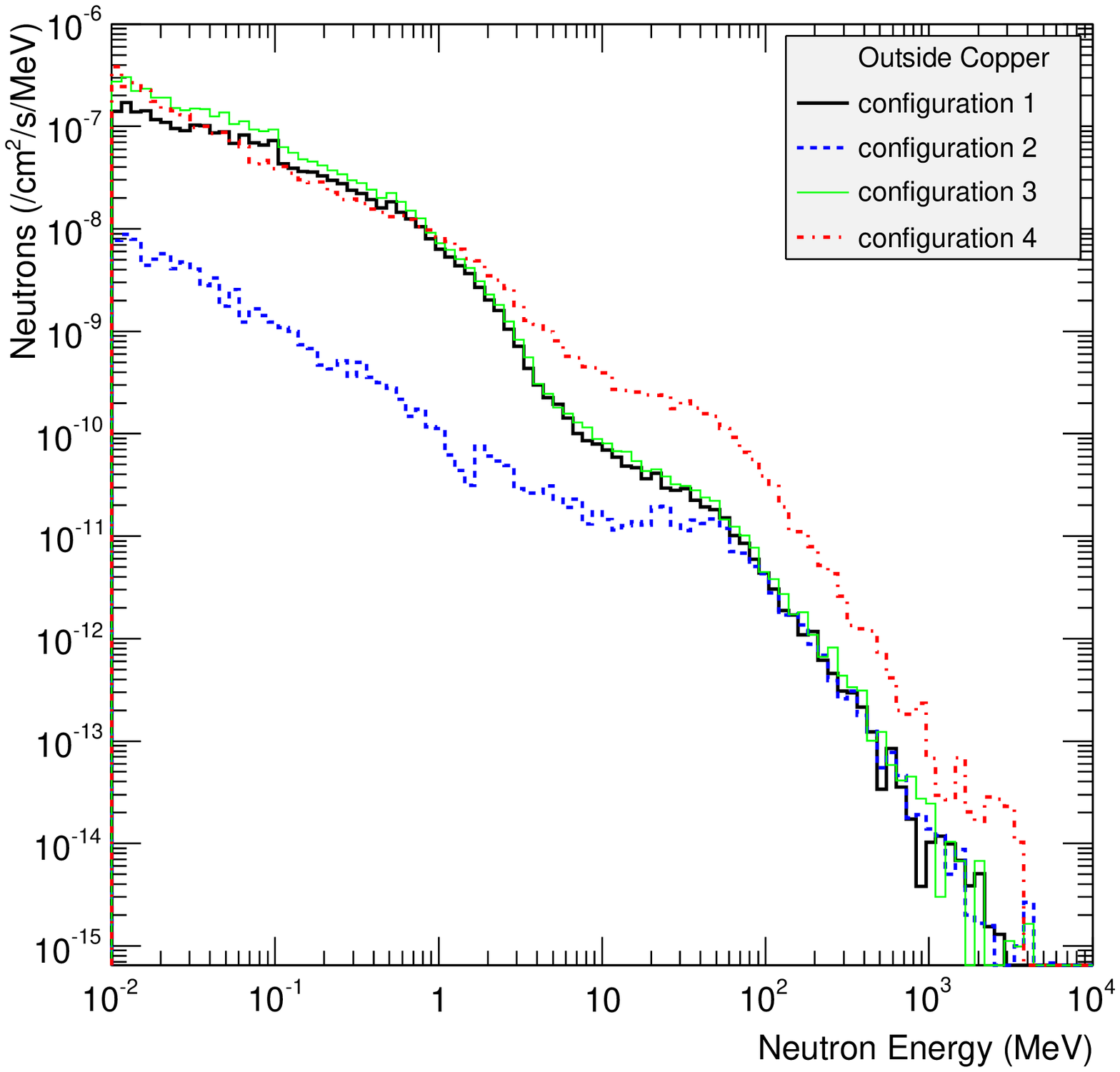}
\includegraphics[width=0.49\columnwidth]{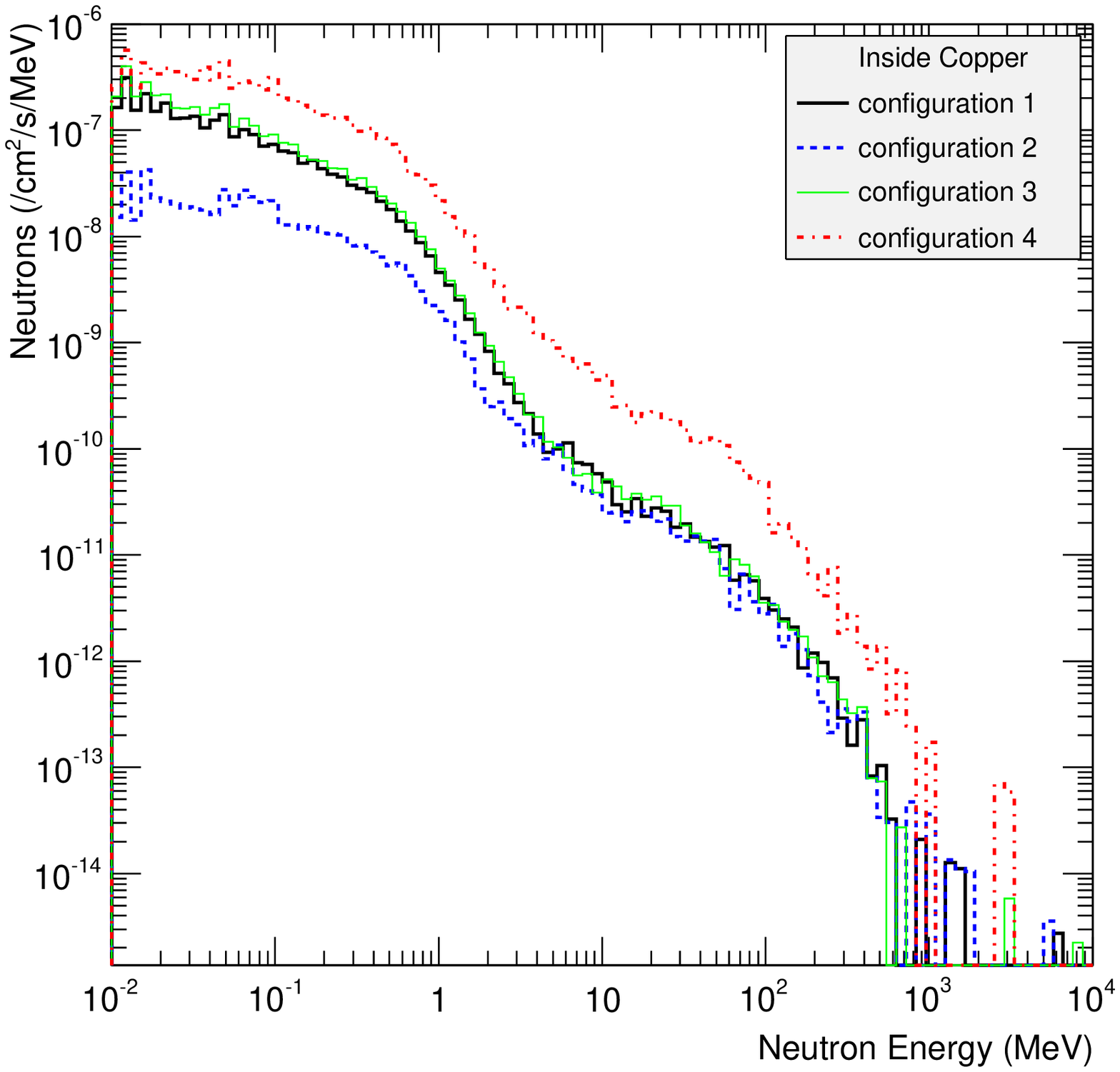}
\caption{Effect of neutron shielding with different configurations.}
\label{shieldingeffect}
\end{center}
\end{figure}

In addition, we tested the shielding effects with respect to shielding configurations
by comparing neutron energy spectra from configurations 1, 2, 3, and 4 at two different boundaries. 
One boundary is outside the copper shield and another is inside the copper shield. 
Results at each boundary are shown in the left and the right of Figure~\ref{shieldingeffect}.
There is little difference between configuration 1 and 3 for neutrons below $\sim$0.1~MeV.
If we consider the order of magnitude difference in muon fluxes at the depths of Gran Sasso and Y2L,
the shielding effect with configuration 2 is very similar to the result with configuration 4, 
which used 5~cm thick PE inside the lead shield and 30~cm thick MD instead of PE in configuration 1.
As shown in the left of Figure~\ref{shieldingeffect}, 
even 5~cm thick PE inside the lead is effective to shield neutrons below $\sim$1~MeV.
Accordingly, 30~cm thick PE inside the lead shield~(configuration 2) is 
the most effective in shielding intermediate neutrons compared to other configurations.

\subsection{Neutron detection in NaI crystals}
From the simulation studies described in sections~\ref{sec3.3} and \ref{sec3.4}, we obtained the neutron flux at the boundaries of the shield layers 
and tested the shielding effect in each shielding material. 
We now consider the nuclear recoil event in NaI crystals due to neutron backgrounds.

\begin{figure}[!htb]
\begin{center}
\includegraphics[width=0.6\columnwidth]{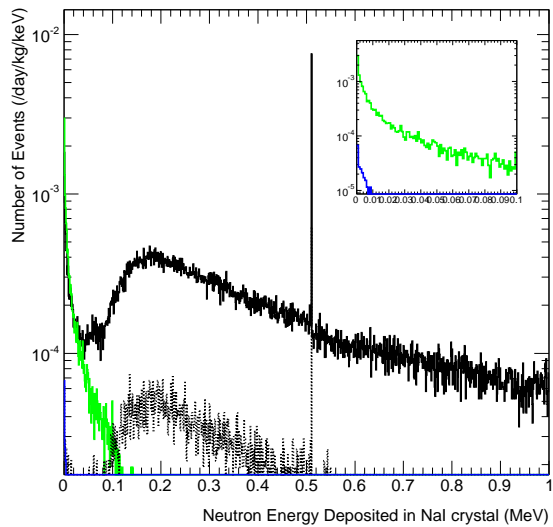}
\caption{Energy spectra deposited in NaI crystals. Results are obtained from the simulation with shielding configuration 1 at Gran Sasso. The thick black line represents all events that deposited energy in the NaI crystal and the dotted black line represents single hit events. The green color indicates energy deposited by all nuclear recoil events, and the blue represents results from nuclear recoil single hit event only.}
\label{singleEdep}
\end{center}
\end{figure}

Because we have 5$\times$5 segmented NaI crystals an event can deposit its energy in two or more crystals. 
Therefore, we call an event a single hit event when it deposits the energy in a single NaI crystal and a multiple hit event 
when it exhibits energy deposition in two or more NaI crystals.
Figure~\ref{singleEdep} shows the spectra of energy deposited in NaI crystals, which include not only nuclear recoil but also electromagnetic energy deposition. 
The thick black line represents all events that deposited energy in the NaI crystal and the dotted black line represents single hit events. 
The green color represents energy deposited by all nuclear recoil events, and the blue color represents results for nuclear recoil single events only. 
The nuclear recoil events contribute only at the lowest energies. 
Figure~\ref{singleEdep} is obtained from the simulation with shielding configuration 1 at Gran Sasso, and for nuclear recoil single hit events it is much lower than the $10^{-4}$ cpd level for energies below 100~keV~(inset).

\begin{figure}[!htb]
\begin{center}
\includegraphics[width=0.49\columnwidth]{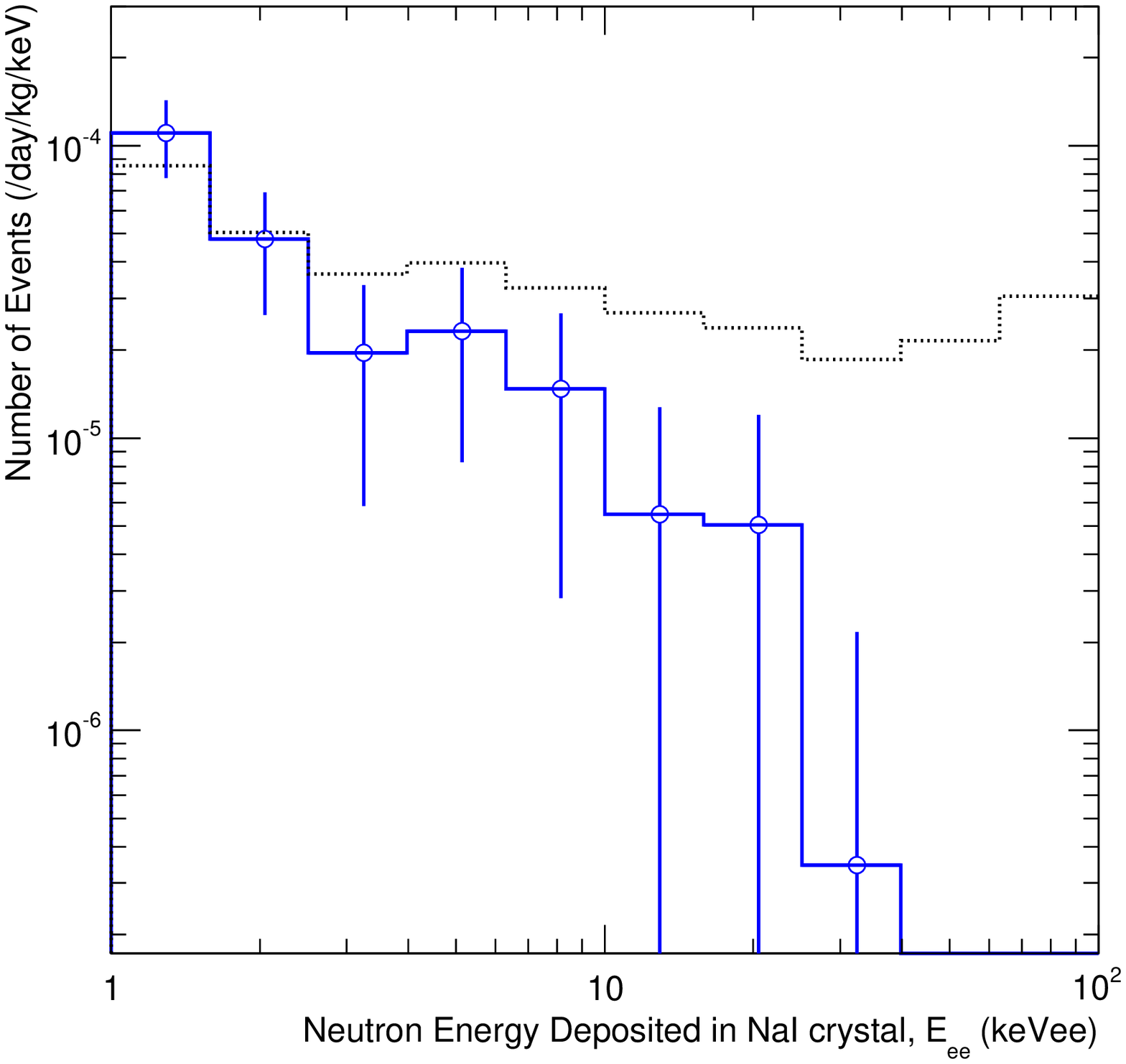}
\includegraphics[width=0.49\columnwidth]{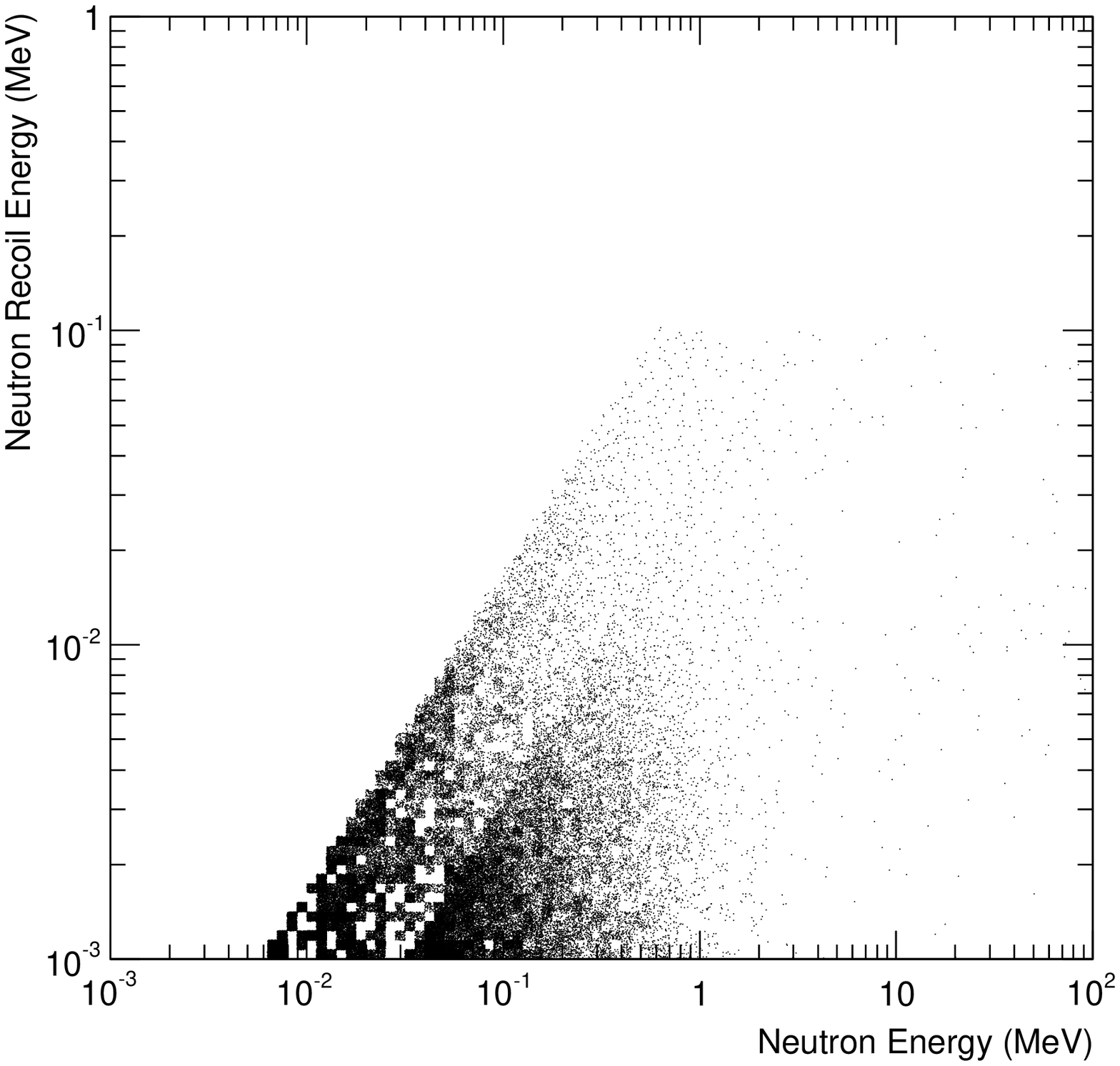}
\caption{Neutron energy deposited in NaI crystals by single hit events~(left). The blue histogram is obtained from nuclear recoil single hit events, and the dotted black histogram results from single hit events, including electromagnetic energy deposition as well as nuclear recoil. 
Quenching factors, 0.25 and 0.1, for Na and I are applied in the simulation.
The correlation between neutron energy and nuclear recoil energy is shown as a scatter plot~(right).}
\label{nuclearRecoil}
\end{center}
\end{figure}

The left of Figure~\ref{nuclearRecoil} shows the spectra of visible energy deposited by all single hit events~(dotted black line), using configuration 1. 
The thick blue line represents results from nuclear recoil single hit events only.
A nuclear recoil event yields a lower number of scintillation photons 
than an electron recoil events for the same energy deposition.
Measurements of the quenching factors for Na and I recoils in NaI(Tl) crystals have been reported and are $\sim$25 and $\sim$10$\%$, respectively~\cite{quench}.
Therefore, we applied 0.25 and 0.1 as quenching factors for Na and I in the simulation.
The right graph in Figure~\ref{nuclearRecoil}  depicts the relationship between nuclear recoil energy and neutron kinetic energy as a scatter plot,
and the Na and I recoils are separated. 
   
\section{Results}
Table~\ref{result} shows the summarized results for single hit event rates for the four shielding configurations. 
Single hit events include electromagnetic energy deposition as well as nuclear recoil energy deposition. 
The single hit event rates by the nuclear recoil event only are also obtained. 

The nuclear recoil single hit event rates at Y2L and Gran Sasso are (4.59$\pm$1.05)$\times10^{-5}$  and (8.69$\pm$1.45)$\times10^{-6}$ dru, respectively, 
and all single hit events rates are on the order of $10^{-4}$ and $10^{-5}$ dru for configurations 4 and 1.

\begin{table}[!htb]
\begin{center}
\caption{Neutron background events in dru unit (events/day/kg/keV) in 5$\times$5 segmented NaI crystals with different shielding configurations at two different underground sites}
\label{result}
\begin{tabular}{c|c|c|c}\hline
& & & Nuclear Recoil  \\ 
& Shielding & Single hit Events & Single hit Events \\
Place & configuration & [/day/kg/keV] & [/day/kg/keV]  \\ \cline{3-4}
&  & Energy [keVee] & Energy [keVee] \\
&  & 2 - 10 & 2 - 10 \\ \hline
GS & 1 & (1.62$\pm$0.20)$\times10^{-5}$ & (8.69$\pm$1.45)$\times10^{-6}$ \\
GS & 2 & (6.52$\pm$1.25)$\times10^{-6}$ & (1.93$\pm$0.68)$\times10^{-6}$ \\
GS & 3 & (1.84$\pm$0.21)$\times10^{-5}$ & (1.35$\pm$0.18)$\times10^{-5}$ \\ 
Y2L & 4 & (1.21$\pm$0.17)$\times10^{-4}$ & (4.59$\pm$1.05)$\times10^{-5}$ \\
\end{tabular}
\end{center}
\end{table}


\section{Conclusion} 
As shown in Table~\ref{result}, it is very unlikely that the neutrons are the main source for the DAMA annual modulation, 
which was reported as (0.0112$\pm$0.0012) cpd/kg/keV in the (2-6)~keV energy interval of the modulation amplitude \cite{bernabei13}.

The DAMA group refuted Davis's claim because the induced modulation
amplitudes from neutrons induced by muons and by neutrinos are less
than  9 $\times 10^{-6}$ cpd/kg/keV (2 $\times 10^{-5}$ cpd/kg/keV for
neutrons produced in the lead shield) and less than 2 $\times 10^{-��6}$
cpd/kg/keV, respectively from the simulation, and they are less than  0.1\% of 
the modulation amplitude measured by DAMA/LIBRA~\cite{bernabei2014}.

\end{document}